\documentclass[amsmath,amssymb,reprint,superscriptaddress]{revtex4-1}
\usepackage{graphicx}
\usepackage{xcolor}
\usepackage[colorlinks=true,urlcolor=blue,citecolor=blue,linkcolor=blue]{hyperref}

\usepackage{physics}
\usepackage{siunitx}
\usepackage{xspace}
\usepackage{bm}
\usepackage{comment}
\usepackage{lineno}

\usepackage[normalem]{ulem}

\begin{document}
\title{Optical non-linearities and spontaneous translational symmetry breaking in driven-dissipative moir\'e exciton-polaritons }
\author{A. Camacho-Guardian}
\affiliation{Departamento de F\'isica Qu\'imica, Instituto de F\'isica, Universidad Nacional Aut\'onoma de M\'exico, Apartado Postal 20-364, Ciudad de M\'exico C.P. 01000, Mexico\looseness=-1}
\author{N. R. Cooper}
\affiliation{T.C.M. Group, Cavendish Laboratory, University of Cambridge, JJ Thomson Avenue, Cambridge CB3 0HE, United Kingdom\looseness=-1}
\affiliation{Department of Physics and Astronomy, University of Florence, Via G. Sansone 1, 50019 Sesto Fiorentino, Italy\looseness=-1}
\date{\today}
\begin{abstract} 
Moir\'e lattices formed from semiconductor bilayers host tightly localised excitons that can simultaneously  couple strongly to light and possess large electric dipole moments. This facilitates the realization of new forms of polaritons that are very strongly interacting and that have been predicted to lead to strong optical nonlinearities controlled by multi-photon resonances. Here, we investigate the role of the non-local component of the exciton-exciton (dipolar) interactions on the optical response of these strongly-interacting moir\'e exciton-polaritons under conditions of strong optical driving. We find that the non-local interactions can strongly influence the steady-state properties leading to multi-stabilities with spontaneously broken translational symmetry and pronounced distortions of the multi-photon resonances. We develop a self-consistent approach to describe the steady-state solution of moir\'e excitons coupled to a  cavity field, treating the long-range interaction between the excitons and the photon field at the semi-classical level.
\end{abstract}
\maketitle
\maketitle
\section{Introduction}

In van der Waals bilayers, the moir\'e superlattice resulting from lattice mismatch or relative twist angle has emerged as a productive means to realize complex quantum many-body phases~\cite{Andrei2021}. The quantum confinement provided by the moir\'e landscape has unfolded many opportunities towards the controlled realization of strongly correlated electronic phases such as Wigner crystals~\cite{Zhang2020}, Mott insulators~\cite{Regan2020}, superconductivity~\cite{Balents2020,Kezilebieke2022} and more~\cite{Cao2018, Yankowitz2019,Serlin2020,Shimazaki2020,Tang2020,Huang2021}. In semiconductor bilayers, moir\'e materials have unveiled a new class of excitations~\cite{Alexeev2019,Jin2019,Tran2019,Seyler2019,Liu2021}: moir\'e excitons. Moir\'e excitons possess properties that make them ideal to explore strongly interacting phases of bosonic matter in uncharted territory, which include  prospects for high-temperature and long-lived Bose-Einstein condensates~\cite{Lagoin2021,Remez2021}, the superfluid-Mott transition~\cite{Gotting2022}, excitonic insulators~\cite{Zhang2021a}, and supersolidity~\cite{Julku2022}.

When combined with an optical cavity, the underlying nature of moir\'e excitons leads to novel forms of exciton-polaritons and presents new opportunities to engineer hybrid quantum states of light and matter with no equivalent in conventional polaritons. The strong confinement of excitons to the moir\'e sites yields  distinctive moir\'e-induced polaritons~\cite{Zhang2021}, novel forms of quantum emitters~\cite{Yu2017,Baek2020,Camacho2021}, and promises a new generation of polaritons with tuneable features~\cite{Yu2020,Fitzgerald2022}. The moir\'e superlattice activates a rich interplay between the tight confinement of the excitons, the light-matter coupling, and the strong exciton-exciton interactions.

 \begin{figure}[h!]
\centering
    \includegraphics[width=\columnwidth]{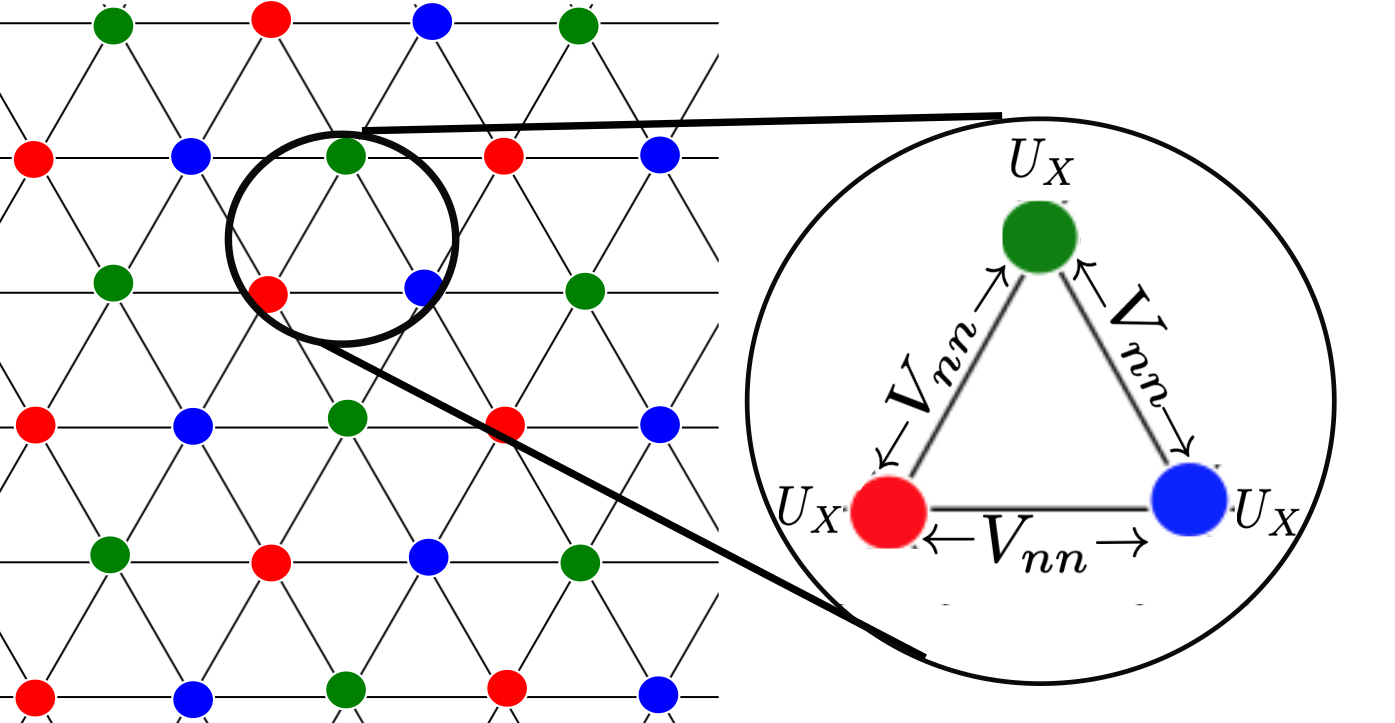}
    \caption{Excitons tightly confined into moir\'e sites forming a triangular lattice. Three distinct exciton sites, here illustrated by the red, green, and blue colours, are treated independently. Excitons interact through an on-site interaction $U_X$ and non-local interaction between excitons in different sites. Here we illustrate the nearest neighbour coupling $V_{nn}$ but dipolar coupling to all distances is included in our calculation.} 
    \label{Fig1}
\end{figure}

Moir\'e polaritons are particularly interesting as the underlying excitons can inherit properties of both spatially direct and indirect excitons, which can provide them with valuable features such as sizeable light-matter coupling~\cite{Alexeev2019,Ruiz2019} and strong exciton-exciton interactions~\cite{Zhang2021}. {Recent experimental~\cite{Zhang2021} and theoretical studies~\cite{Camacho2021} have demonstrated that moir\'e polaritons feature optical properties with large non-linearities very different from conventional polaritons in semi-conductors. Theoretically, it has been shown that the tight confinement of the excitons to the moir\'e sites leads to pronounced multi-photon resonances governed by the underlying {\it discrete} excitonic energy spectrum~\cite{Camacho2021}, arising as a consequence of the quasi zero-dimensional character of the excitons and their strong local interactions. This is predicted to permit lasing based on single- and multi-photon processes induced by the moir\'e lattice. 

In addition to the strong on-site exciton-exciton interactions, the indirect character of the moir\'e excitons leads to non-local interactions. While for moir\'e lattices the on-site interaction is expected to dominate over the non-local interactions, the precise role of the non-local interactions remains relatively unexplored. The role of non-local interactions on out-of-equilibrium polaritons is further motivated by theoretical predictions and breakthrough experiments where non-local interactions are a key element to stabilizing states with spontaneously broken translational symmetry which can lead to complex many-body phases such as supersolids~\cite{Otterlo1994,Batrouni1995,Scalettar1995,Goral2002}, and which have been already been observed in dipolar quantum gases~\cite{Tanzi2019,Norcia2021,Tanzi2021}. The dipole-dipole interaction between excitons beyond the local interaction has very recently suggested the existence of supersolid phases of moiré excitons~\cite{Julku2022}, and instigates the study of the interplay between these phases and polariton physics.

Here, motivated by this open question, we study the many-body optical properties of a van der Waals heterostructure bilayer, focusing on the effects of the non-local exciton-exciton interactions arising from their dipolar character. We show that non-local interactions can strongly modify the optical response of the system and demonstrate the emergence of steady states with broken translational symmetry. In addition, the presence of non-local interactions influences the multi-photon resonance conditions leading to a rich phase-diagram with strongly hysteretic features.

The outline of the paper is as follows. In Section~\ref{Model}, we detail the model we study -- a tight binding model of excitons coupled to cavity photons -- and the methods we employ to determine its properties. Here, we introduce three coloured excitonic sites which we treat independently at the mean-field level, we also discuss the mean-field and semi-classical treatment for the cavity photons. In Section~\ref{hardcore} we turn our attention to the study of hard-core excitons, and we reveal the emergence of steady states with broken translational symmetry that can be accessed through several hysteresis mechanisms. The interplay between the on-site and non-local interactions is unraveled in Section~\ref{multip} where we analyze the effects of the dipolar interactions have on the multi-photon resonances. Finally, in Section~\ref{conclusions} we discuss the experimental consequences and outlook based on our results.

\section{Model and Methods}
\label{Model}
We consider moir\'e excitons in a van der Waals hetero-structure bilayer coupled to a microcavity in the presence of a coherent drive of photons. The moir\'e landscape leads to flat mini bands that arise from the tight localization of the excitons to the moir\'e sites. Hence, we describe the excitons via a tight binding Hamiltonian given by 
\begin{gather}
\label{HX}
\hat H_X=\sum_{i}\omega_X\hat x^\dagger_i\hat x_i+\frac{U_X}{2}\sum_i \hat x^\dagger_i\hat x^\dagger_i\hat x_i\hat x_i+\sum_{i\neq j}\frac{V_{ij}}{2}\hat x^\dagger_i\hat x^\dagger_j\hat x_j\hat x_i, 
\end{gather}
here $\hat x^\dagger_i$ creates an exciton with energy $\omega_X$ in the site $i$, with $N_s$ sites arranged on a triangular lattice. (We set $\hbar=1$ throughout.) We neglect the hopping of excitons between local sites -- i.e. the bandwidth of the lowest energy exciton band in the moir{\' e} lattice. For a wide range of parameters this can be small compared to the transport via the cavity mode. The on-site exciton-exciton interaction is denoted by $U_X,$ while $V_{ij}$ corresponds to the interaction between an exciton in site $i$ and an exciton in site $j$.  In general, the moir\'e potential supports multiple localised exciton states~\cite{Tran2019,Fitzgerald2022}. Here we restrict our study to the lowest excitonic state and assume that the energy separation between the first and second bands remains larger than any other typical energy of the system.

The bilayer is embedded in a high-finesse microcavity with the ideal dispersion of the cavity photons described by  
\begin{gather}
\hat H_{\rm l}=\sum_{\mathbf k}\omega_c(\mathbf k)\hat a^\dagger_{\mathbf k}\hat a_{\mathbf k}\,.
\end{gather}
Here the free dispersion of photon is $\omega_c(\mathbf k)=\omega_c+|\mathbf k|^2/(2m_c),$ where $m_c$ is the cavity photon mass. The operator $\hat a^\dagger_{\mathbf k}$ creates a cavity photon with in-plane momentum $\mathbf k.$ 
The coupling between excitons and cavity photons is given by the usual light-matter Hamiltonian
\begin{gather}
\hat H_{\rm l-m}=\sum_{\mathbf k}\Omega\left(\hat a_{\mathbf k}^\dagger \hat x_{\mathbf k}+\hat x^\dagger_{\mathbf k}\hat a_{\mathbf k}\right),
\end{gather}
where $\hat{x}^\dagger_{\mathbf k}$ creates an exciton with in-plane momentum $\mathbf k.$ The strength of the light-matter coupling, denoted by the Rabi frequency $\Omega$, is assumed much smaller than the typical energy  of the excitons, so the light-matter Hamiltonian is written under the rotating wave approximation. (For typical systems $\Omega$ is a several meV while $\omega_X$ is on the order of  1 eV.) We, however, consider the regime of strong light-matter coupling such that the Rabi coupling is much larger than the cavity losses $\gamma_c,$ and light couples efficiently to the excitons. 

To make further progress we shall restrict the light-matter coupling to the $\mathbf k=0$ cavity mode~\cite{Camacho2021}. Thus, we simplify the light and light-matter terms of the Hamiltonian as
\begin{gather}
\hat H_{\rm l} + \hat H_{\rm l-m} = \omega_c \hat{a}^\dag \hat{a} + \frac{1}{\sqrt{N_s}}\sum_{ i}\Omega\left(\hat a^\dagger \hat x_{i}+\hat a \hat x_{i}^\dagger \right)\,,
\end{gather}
where $\hat{a}$ now refers to the $\mathbf k=0$ cavity mode. 
This approximation assumes a spatially uniform coupling between the cavity photons and excitons occupying different sites in the moir\'e superlattice. This assumption is justified for two reasons. Firstly, due to the ultra-light mass of the cavity mode, the cavity photons decouple from the excitons when the kinetic energy of the photon becomes of the order of the Rabi coupling $k^2/2m_c=\Omega$. This means that excitons only couple to cavity modes with wavelengths that are larger  than a length scale $\lambda=2\pi/k$, which is typically very large, covering hundreds of moir\'e sites. This large lengthscale justifies a mean-field treatment of the cavity-mediated exciton-exciton coupling. Secondly, we consider situations in which the cavity mode is pumped uniformly. Specifically, we consider an external coherent injection of photons via
\begin{gather}
\hat H_{\text{drive}}=(F\hat a^\dagger e^{-i\omega_p t}+F^*\hat a e^{i\omega_p t}),
\end{gather}
where $F$ and $\omega_p$ are the strength and the frequency of the driving term respectively. As will be discussed below, we will employ a mean-field approach for the cavity photons, where excitons couple only to the $\mathbf k=0$ cavity mode. We describe the system in the rotating frame of this light field, using $\hat{\tilde a}=\hat a e^{i\omega_p t}$ and $\hat{\tilde x}_{\mathbf k}=\hat x_{\mathbf k} e^{i\omega_p t},$ for simplicity we drop the $\sim$'s  in the following. The rotating frame introduces the pump energy detuning from the exciton and cavity detuning  defined as $\Delta\omega_X=\omega_p-\omega_X$ and $\Delta\omega_c=\omega_p-\omega_c,$ respectively. 

 We mention that the ability to create arbitrary number of excitons is limited by the intrinsic nature of the excitons, where its non-bosonic nature leads to saturation effects. Although this remains as an open question~\cite{Combescot2007,Levinsen2019}, such effects can  be accounted to a first approximation through an anharmonic light-matter coupling term~\cite{Ciuti2000a,Schmitt1985,Baas2004}. This term, however, only quantitatively modifies the optical properties for small exciton numbers~\cite{Camacho2021}.

We allow the system to be lossy, and study the density operator of the system, $\hat \rho$, via the Gorini–Kossakowski–Sudarshan–Lindblad master equation 
~\cite{Carusotto2013}
\begin{gather}
\frac{d\hat\rho}{dt}=-i[\hat H,\hat\rho]+\mathcal D[\hat\rho]=\mathcal L[\hat\rho]\,,
\label{ME}
\end{gather}
where the total Hamiltonian is given by $\hat H=\hat H_X+\hat H_{l}+\hat H_{\rm l-m}+\hat H_{\text{drive}}.$ 
  The dissipative character of the system is accounted for by the  operator 
  \begin{gather}
\mathcal D[\hat\rho]=\frac{\gamma_c}{2}\left[2\hat a\hat\rho\hat a^\dagger-\{\hat a^\dagger\hat a,\hat\rho\}\right]+\sum_{i}\frac{\gamma_x}{2}\left[2\hat x_i\hat\rho\hat x^\dagger_i-\{\hat x^\dagger_i\hat x_i,\hat\rho\}\right],
\label{eq:excitonandcavitydecay}
\end{gather}
here $\gamma_x$ and $\gamma_c$ are the damping rate of the excitons and photons respectively. 

The spatial stacking of the monolayers leads to a moir\'e periodicity that confines the excitons into a triangular lattice. To allow spatial ordering of the excitons we introduce a 3-site supercell as illustrated in Fig.~\ref{Fig1}. We thus define three kinds of sites within a larger unit cell, denoted in the figure (and referred to below) by three different colours: red, green, and blue. By treating these three sites independently, our present study extends beyond Ref.~\cite{Camacho2021} to explicitly permit states with broken translational symmetry. We will show that this can arise as a consequence of the  non-local interaction $V_{nn}.$

 We employ a mean-field approximation to decouple the dipolar interactions in Eq.~\ref{HX}. We define a supercell as illustrated in Fig.~\ref{Fig1}, treating the occupations of all sites of the same colour to be the same. We thus replace $\hat{x}_{i}\to \hat{x}_{\alpha}$ where $\alpha \in (\text{R,G,B})$ labels the three distinct sites within the supercell. The Hamiltonian for each colour is then  
\begin{gather}
\label{EqHX}
\hat H_{X,\alpha}=\left(\omega_X+{V_{\rm d}}n_\alpha +{V_{\rm od}}\sum_{\beta\neq\alpha} n_{\beta}\right) \hat x^{\dagger}_{\alpha}\hat x_{\alpha}+\\ \nonumber
+\frac{U_X}{2} \hat x^\dagger_{\alpha}\hat x^\dagger_{\alpha}\hat x_{\alpha}\hat x_{\alpha},
\end{gather}
this local Hamiltonian treats the local terms in Eq.~\ref{HX} exactly, but takes the long-range dipolar interaction at the mean-field level, with the expectation values of the exciton number for the three different sites, $n_\beta$, to be determined self-consistently. Here, $V_{\rm d} = 2.124\times V_{nn}$ and $V_{\rm od} = 4.455 \times V_{nn}$ give the mean-field dipole-dipole interaction between the site $\alpha$ in the supercell and all of the different sites with same and different colour respectively, see further details in the Appendix~\ref{AppendixN} and Ref.~\cite{Lambin1993}. Here, $V_{nn}=d^2/a_M^3$ is the dipole-dipole interaction between excitons in nearest neighbour sites, where $a_M$ is the moir\'e period of the superlattice and $d$ is the dipole moment of the hybrid exciton due to its interlayer charge separation.  

On the other hand, the light-matter coupling is insensitive to our artificial distinction of sites, that is, cavity photons couple equally to excitons regardless of the colour of the site they occupy. The light-matter coupling for a given colour simply reads as 
\begin{gather}
    \hat H_{\rm l-m,\alpha}=\frac{\Omega}{\sqrt{N_s}}\left(\hat a^\dagger \hat x_{\alpha}+\hat a \hat x_{\alpha}^\dagger\right).
\end{gather}

To make further progress, we take a semi-classical approach for the cavity photons, where we replace the cavity field by its expectation value $\langle \hat a\rangle=\sqrt{N_s}\psi_a$. In this case, the steady-state solution for the photon amplitude $\psi_a$ is given by
\begin{gather}
\label{psi}
\psi_a=\frac{1}{\Delta_c}\left(f+\frac{\Omega}{3}\sum_{\alpha=\text{R,G,B}}\langle \hat x_\alpha\rangle\right),
\end{gather}
where $f=F/\sqrt{N_s}.$ The last term inside of the brackets accounts for the possible different expectation value of sites with different colour. Here, $\Delta_c=\Delta\omega_c+i\gamma_c/2.$ 

These approximations permit us to define three Hamiltonians for differently colours $\alpha$,
\begin{gather}
\hat H_{\text{loc},\alpha}=\hat H_{X,\alpha}+\hat H_{\rm l-m,\alpha},
\end{gather}
with $\alpha,\beta\in (\text{R,G,B}),$  that are local. 
The coloured Hamiltonians are coupled through the exciton-exciton interactions and the light-matter coupling. The exciton-exciton interaction couples sites with different and same colours via the terms $V_{\text{od}},$ and $V_{\text{d}},$ in Eq.~\ref{EqHX} treated at the mean-field level. This mean-field approximation can be understood intuitively: it introduces a self-consistent on-site energy that can vary for the three different colours and thus, can energetically favour the breaking of the translational symmetry.  That is, the imbalanced-population steady-state solutions are a consequence of an emergent colour-dependent on-site energy arising from the non-local interactions. We emphasize that since our starting Hamiltonian has the full translational symmetry of the triangular lattice, the emergence of collective phases with a reduced translational symmetry through interactions is an example of spontaneously broken symmetry~\cite{AltlandSimons}. Thus there is a residual (discrete) degeneracy of the ground state associated with the different ways in which these broken symmetry states can be placed on the triangular lattice. Which of the broken-symmetry states appears in the numerics is, as usual, determined by the initial seed.
On the other hand, the light-matter coupling introduces a long-range mediated tunneling, where an exciton in a given site can convert to a cavity photon, which can decay into an exciton in any other moir\'e site. Thus, the long-range photon-mediated hopping couples sites with the same and different colours. This leads to a coupling between the coloured local Hamiltonians. That is, we assume that the cavity field retains the spatially uniform coupling to the excitons in the presence of non-local interactions and that the cavity field maintains population only in the $\mathbf k=0$ mode\cite{Camacho2021}.

Our approach leads to a set of three coupled master equations, for the sites of each different colour
\begin{eqnarray}
\frac{d\hat{\rho}_\alpha}{dt} & \equiv & {\cal L}_\alpha [\hat{\rho}_\alpha] \\ \nonumber
 & = & -i[\hat{H}_{{\rm loc},\alpha},\hat{\rho}_\alpha] +\frac{\gamma_{x}}{2}
\left[2\hat x_\alpha \hat\rho_\alpha \hat x_\alpha^\dagger-\{\hat x^\dagger_\alpha\hat x_\alpha,\hat\rho_\alpha\}\right],
\end{eqnarray}
To obtain the steady-state properties we employ exact diagonalization of each of these three equations.
They are coupled since, via Eqn.(\ref{psi}), the expectation values of the photon amplitude  $\psi_a$ and the exciton number $\langle x^{\dagger}_{\beta}\hat x_{\beta}\rangle$ must be obtained self-consistently. The numerical scheme is detailed in the Appendix~\ref{appendixA}.

The long-range interaction between the excitons stems from their indirect nature that leads to a dipole-dipole interaction that scales with separation $r$ as $1/r^3.$ Due to the large moir\'e periodicity, the on-site interaction $U_{X}$ is largely dominant with respect to $V_{nn}.$ While the on-site interaction can be made of the order of some tens meV~\cite{Zhang2021}, the interaction between first neighbours is estimated $V_{nn}\sim 0.1-1 \text{meV}$~\cite{Julku2022}, thus, for typical experiments one expects $V_{nn}/U_X\approx 10^{-1}-10^{-2},$ which enables the study of the imprints of the non-local interactions over a wide range of parameters. We expect our approximation to be valid when the non-local interaction remains smaller compared to the on-site interaction $V_{\text{d}}/U_X\ll 1$ and $V_{\text{od}}/U_X\ll 1.$ We consider two cases, first, we consider hard-core excitons which are prevented from double occupation. Then, we study the interplay between multi-photon resonances and non-local interactions.

\section{Hard-core excitons}
\label{hardcore}

For clarity, we start our study in the limit of hard-core excitons which explicitly forbids multiple occupation, this case will allow us to understand the effects of the long-range interaction disentangled from the on-site interaction. 

For hard-core excitons $U_X\to\infty$ non-linearities arise from the impossibility to create multiple excitons per site and from the non-local interactions. First, we explore steady-state solutions with equal population on the three sites. We start by considering an initial seed for our self-consistent scheme that is population balanced  ($n_{\rm R}=n_{\rm G}=n_{\rm B}$), see Appendix~\ref{appendixA}.  In this case, we find a pair of solutions corresponding to low- and high-density hysteresis branches. The former corresponds to varying  $f/V_{nn}$ from below while the latter arises from  $f/V_{nn}$ being tuned from above. These solutions are illustrated in Fig.~\ref{Fig2} for $\Omega/V_{nn}=0.5,$ while fixing the losses $\gamma_x/V_{nn}=\gamma_c/V_{nn}=0.1$ and cavity and exciton detunings of $\Delta\omega_c/V_{nn}=-1$ and $\Delta\omega_X/V_{nn}=0.9,$ respectively.

 \begin{figure}[h!]
\centering
    \includegraphics[width=\columnwidth]{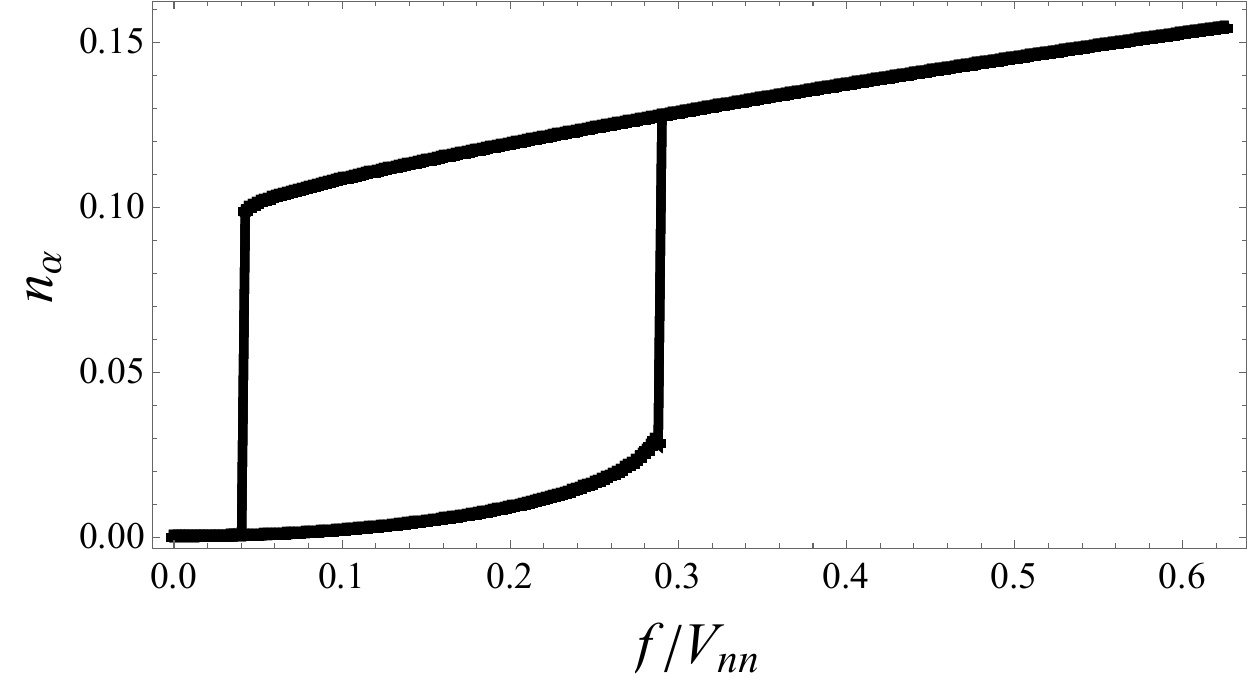}
    \caption{Exciton number per site for the low- and high-density branches of the hysteretic response, in the hard-core limit $U_X\to\infty$. Here we consider only uniform states in which $n_\alpha$ is independent of site $\alpha$.  We take $\Omega/V_{nn}=0.5,$ with $\gamma_x/V_{nn}=\gamma_c/V_{nn}=0.1,$ and the detunings are $\Delta\omega_c/V_{nn}=-1$ and $\Delta\omega_X/V_{nn}=0.9.$
    } 
    \label{Fig2}
\end{figure}

The behaviour just described appears qualitatively similar to that found in Ref.~\cite{Camacho2021}, where it was shown that strong on-site interactions and the inherent nature of moir\'e exciton-polaritons can give rise to new physical phenomena absent in conventional polaritons~\cite{Zhang2021,Camacho2021}. Fig.~\ref{Fig2} illustrates that the transition from a low-density regime ($f/V_{nn}\lesssim 0.05$) to a regime dominated by the driving ($f/V_{nn}\gtrsim$ 0.3) is separated by an intermediate regime where exciton-exciton interactions lead to an apparent  bistability.

Intriguingly, as we now show,  in this regime we find that the presence of non-local interactions can lead to additional steady-state solutions, turning the bistability into a multi-stability with four steady-states: the two solutions in Eq.~\ref{Fig2} and two additional solutions with spatial ordering of the excitons, which we will now discuss. These solutions are illustrated in Fig.~\ref{Fig3}  for the same values of the parameters as in Fig.~\ref{Fig2}, 

The ranges of values of the drive corresponding to states with  broken translational symmetry are shaded in pink. In these regimes the different sites (labelled by different colours) have unbalanced populations.

\begin{figure}[h!]
\centering
    \includegraphics[width=\columnwidth]{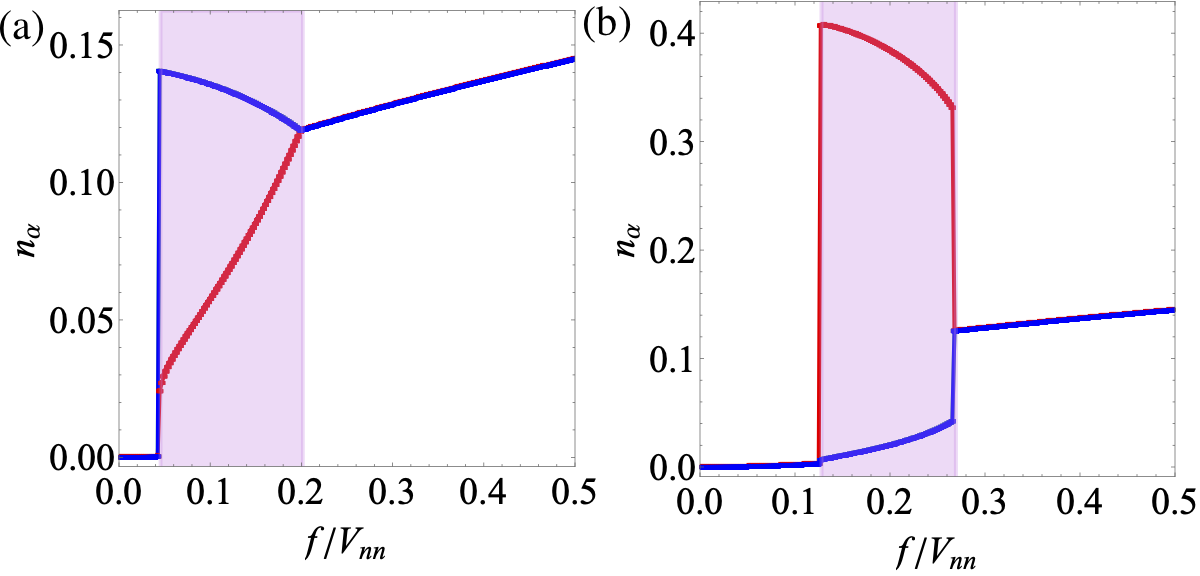}
    \caption{Exciton number per site. (a) Steady-state solutions with $n_{\rm B}=n_{\rm G} > n_{\rm R}$, i.e. one site of low density. (b) Steady-state solutions exhibiting a regime with $n_{\rm R}> n_{\rm B}=n_{\rm G}$, i.e. one site of high density. The pink area illustrates the regime with unbalanced populations. Color code follows the lattice colouring. Parameters are the same as for Fig.~\ref{Fig2}. 
    } 
    \label{Fig3}
\end{figure}

Figure~\ref{Fig3}(a) shows steady-state solutions obtained via hysteresis where a regime with spontaneously broken translational symmetry emerges, characterised by $n_{\rm B}=n_{\rm G}\neq n_{\rm R}$. To reach these steady states, the initial seed of our self-consistent scheme is required to have the same broken translational symmetry as the final state. Therefore, we consider an initial seed that retains hysteretically the solutions for the blue and green sites while sets to zero the density and coherence of the red sites, see further details in Appendix~\ref{appendixA}. In this regime (pink area), two solutions remain in a high-density phase below the bifurcation point, whereas the third colour slowly transits to low densities. Figure.~\ref{Fig3}(a) illustrates the case where blue and green moir\'e sites are in a high density phase and the red tends to low densities. However, one can find the equivalent solutions with  $n_{\rm B}=n_{\rm R}\neq n_{\rm G}$ and  $n_{\rm R}=n_{\rm G}\neq n_{\rm B}$ that exhibit identical features (not shown).

Besides, additional solutions arise  where only one coloured site suddenly jumps into a high-density phase at expenses of two moir\'e sites less populated. These solutions are illustrated in Fig.~\ref{Fig3}(b) where the red sites prevail in a high-density phases with much larger population than the blue and green exciton sites, here $n_{\rm R}\neq n_{\rm B}= n_{\rm G}.$ These steady-state solutions are accessed with a seed where the initial self-consistent parameters for the blue and green sites are set to zero while for the red sites the high-density hysteresis branch is followed, see details in Appendix~\ref{appendixA}. {We have explored different self-consistent schemes, allowing for {\it fully} population-imbalanced metastable states with $n_{\rm R}\neq n_{\rm B}, $   $n_{\rm B}\neq n_{\rm G}, $ and $n_{\rm R}\neq n_{\rm G}$, but our calculations show that these do not appear over the wide range of parameters that we have explored. We speculate that the absence of these solutions arises from the fact that each site can transit either to a low- or high-density phase, this binary characterisation of the population of the sites yields to only four kinds of different solutions shown in Fig.~\ref{Fig2} and Fig.~\ref{Fig3}. }

The photon amplitude in Eq.~\ref{psi}, which couples collectively to the sites with different colours, can also be used as a witness of this  set of solutions with broken translational symmetry for the excitons. This is illustrated in Fig.~\ref{Fig8} where we show the photon number $n_p=|\psi_\alpha|^2$ for the various hysteresis branches discussed above. Since for excitons the low to high density transition occurs for different values of $f,$ the photon number also exhibits this strong hysteresis dependence. Figure~\ref{Fig8}(b) shows the photon number for the same parameters as Figs.~\ref{Fig2}-\ref{Fig3}.  Here, the black lines correspond to the photon densities  when the excitons are uniformly distributed in the moir\'e lattice, that is, the population-balanced low- and high-density branches shown in Fig.~\ref{Fig2}. The red lines illustrate the photon densities of steady-state solutions for which the exciton occupations exhibit broken translational symmetry as in Fig.~\ref{Fig3}(a)-(b). The pink area corresponds to the regime where a steady state can be found with broken translational symmetry.

 \begin{figure}[h!]
\centering
    \includegraphics[width=\columnwidth]{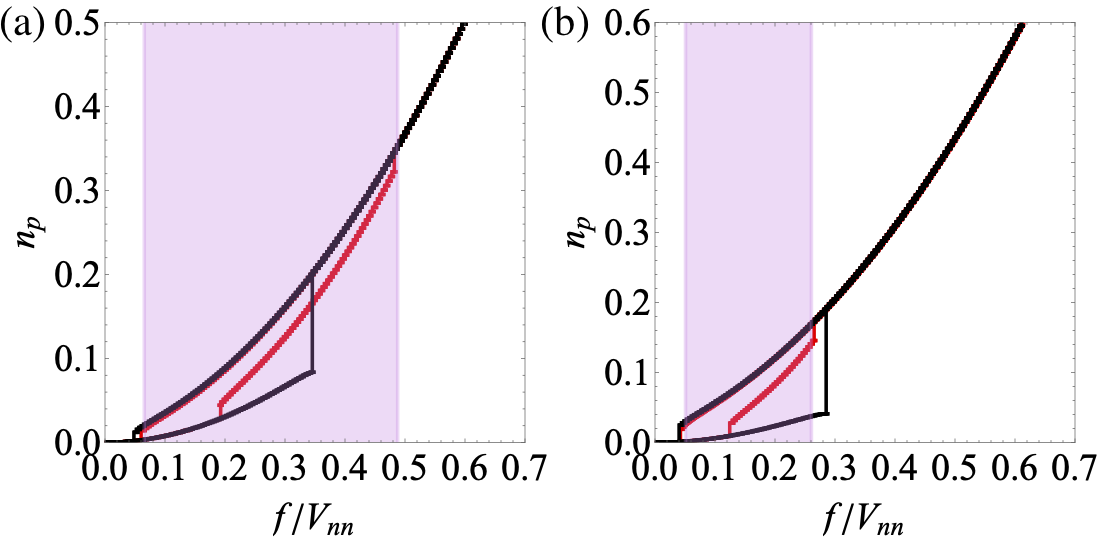}
    \caption{Photon amplitude for the several hysteresis mechanism. (a) $n_p$ for $\Omega/V_{nn}=0.35$ and (b) $\Omega/V_{nn}=0.5$ and remaining parameters as in Figs.~\ref{Fig2} and \ref{Fig3}. Black lines correspond to steady-state solutions with  population balanced exciton number. The photon numbers for steady states with spatial ordering of excitons are illustrated by the red curves. The pink area illustrates the regime where excitons can have density order.
        } 
    \label{Fig8}
\end{figure}

The size of the pink area is determined by the ratio between the light-matter coupling and the strength of the non-local interactions. To illustrate this point we show  in Fig.~\ref{Fig8} the shrinking of these solutions when $\Omega/V_{nn}$ is increased, from $\Omega/V_{nn}=0.35$  to $0.5$. The pink region  clearly decreases for larger values of $\Omega/V_{nn}.$ This illustrates that the solutions with broken translational symmetry are suppressed as for large $\Omega/V_{nn},$ and disappear for $\Omega/V_{nn}>0.65.$   The broadening of the excitonic lines leads to smoothing of the interaction effects, in turn, the large  moir\'e periodicity which gives a small $V_{nn}$ imposes narrow excitonic lines. Experimentally, small broadening of the exciton linewidths have been reported in moir\'e setups, which can be of the order of $0.1\text{meV}$~\cite{Jin2019}.

Thus we have shown that non-local interactions allow for steady-state solutions with spontaneously broken translational symmetry, that is, with unbalanced population in terms of the exciton colour. Numerically, we access states with broken translational symmetry by initializing our self-consisting approach with seeds that explicitly break this symmetry, see further details in the Appendix~\ref{appendixA}. We stress that although the ability to access states with broken translational symmetry depends on the initial seed, the existence of such steady-states hinges on the non-local interactions and as these states are absent when $\Omega/V_{nn}\gg 1$. This dependence on the initial seed is a feature commonly shared for any broken symmetry phenomenon, where the initial seed is used only to select which of the various broken symmetry states is realized. For example, in typical self-consistent schemes for equilibrium Bose-Hubbard-like models, the detection of states with broken translational symmetry requires an initial ansatz that explicitly breaks this symmetry; then, the self-consistent approach can drive the solution into equilibrium states that may preserve the broken translational symmetry. 

Our ansatz is based on the uses of a supercell containing three kinds of moir\'e sites, thus, the solutions follow the restriction imposed by that ansatz. Solutions with different spatial structures, not allowed by our ansatz, could emerge through the use of other supercells. We have also studied another simple two-site supercell which allows for striped density order. We find that striped solutions can also be appeared, but that these are less stable than the structures that we present here. In particular, the striped phases do not appear for the set of parameters discussed in Figs.~\ref{Fig1}-\ref{Fig3}. We cannot rule out that other more complex spatial structures, which are possible only in larger supercells, could be more stable than those we present for the 3-site cell. Also, we expect that disorder could play a significant role in determining the nature of the stable states. For clarity of our presentation, we leave the comprehensive study of more complex broken symmetry phases and disorder to future investigations.

\section{Multi-photon resonances and non-local interactions}
\label{multip}
Now, we turn our attention to the study of the interplay between the on-site interactions and the non-local interactions. Thus, we relax the hard-core constraint and allow for multiple occupation. 
In absence of non-local interactions, that is, for $V_{nn}=0,$ the phase-diagram is governed by the multi-photon resonance condition
\begin{gather}
\label{Nres}
N\omega_p=N\omega_X+\frac{U_X}{2}N(N-1),    
\end{gather}
which leads to the condition 
\begin{gather}
\label{Nres-as-ratio}
    \frac{2\Delta \omega_X}{U_X} = (N-1)\,.
\end{gather} 
Physically this can be understood in terms of an energetic condition dictating that $N$ exciton resonances are promoted whenever the energy of $N$ non-interacting photons  matches the energy of $N$ interacting excitons~\cite{Camacho2021}. 
 \begin{figure}[ht!]
\centering
    \includegraphics[width=1\columnwidth]{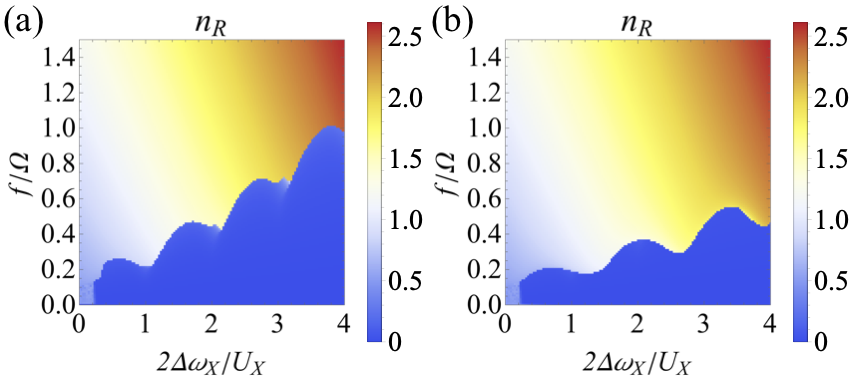}
    \caption{Exciton number per site for steady states with balanced colour population. (a) Low-density hysteresis branch and (b) high-density hysteresis branch. We take   $V_{od}/U_X\approx 0.1,$ $\Omega/U_X=0.65$ , $\Delta\omega_c/U_X=-1$ and $\gamma_c/U_X=0.2.$} 
    \label{Fig5}
\end{figure}
In the presence of non-local interactions, we expect a shift of this energetic condition: treating $V_{nn}$ at the mean-field level, the interaction between adjacent excitons simply displaces the on-site energy $$
\omega'_X=\omega_X+\left({V_{\rm d}}n_\alpha +{V_{\rm od}}\sum_{\beta\neq\alpha} n_{\beta}\right),$$
thus, physically, one anticipates that for a site with a given colour $\alpha,$ the resonance is displaced to
\begin{gather}
\label{Nresalpha}
N\omega_p=N\left(\omega_X+V_{\rm d}n_\alpha +V_{\rm od}\sum_{\beta\neq\alpha} n_{\beta}\right)+\frac{U_X}{2}N(N-1).
\end{gather}
In this case, the multi-photon resonance (\ref{Nres-as-ratio}) depends on $n_\beta$ and is not longer necessarily  an integer.  One should note that in contrast to the Bose-Hubbard model in equilibrium where the occupation number per site of the insulating phase is pinned to integer values, for driven-dissipative systems the discrete lobular pattern determined by the multi-photon resonance condition is a remnant of the discreteness of the equilibrium Hubbard energy spectrum. However, the occupation number is no longer strictly an integer. That is,  while the modulation of the phase diagram for $V_{nn}=0$ follows very closely the discrete equation in Eq.~\ref{Nres}, the exact value of the exciton number slightly above the low-to-high density transition is not necessarily an integer.

From our earlier analysis on the effects of the non-local interactions we also anticipate the emergence of multi-stabilities. Note that, although the cavity field and $V_{nn}$ are treated at the mean-field level, we still perform a full quantum calculation for the driven excitons on a single site.

We begin by discussing the case of large ratio $\Omega/V_{nn}$, which as explained above, tends to inhibit solutions with spatial ordering. We take  $\Omega/U_X=0.65$ which corresponds to  $\Omega/V_{nn}=1.42$, finally, we consider $V_{\text{od}}/U_X=0.1.$     In Fig.~\ref{Fig5} we show the solutions found for a finite on-site interaction. We turn our attention first to population-balanced solutions obtained with initial seeds that do not break translational symmetry ($n_{\rm R}=n_{\rm B}=n_{\rm G}$) and find two solutions corresponding to the low-density and high-density hysteresis branches.

Figure.~\ref{Fig5}(a) corresponds to the phase-diagram following the low-density branch, that is, $f/U_X$ being tuned from below. In this case, the phase-diagram shows sharp cusp-like features at  detunings closely governed by the {\it bare} multi-photon resonance in Eqs.~\ref{Nres},~\ref{Nres-as-ratio}. Figure.~\ref{Fig5}(a) corresponds to the case where the low to high density transition is promoted from below, that is, it corresponds to $n_\beta\approx 0$ in Eq.~\ref{Nresalpha}, and therefore the cusp-like features are barely shifted away from the resonance condition Eq.~\ref{Nres}. The non-local interaction does introduce some blurring of the cusps, but the locations remain closely tied to  Eq.~\ref{Nres-as-ratio}.

On the high-density hysteresis branch, on the other hand, the transition is crossed from above. In this case, the exciton number of the adjacent sites is relatively large $n_\beta\neq 0,$ hence, the multi-photon resonances in Eq.~\ref{Nresalpha} acquire large energy shifts away from Eq.~\ref{Nres-as-ratio} and visibly distort the phase-diagram. The deviations of the multi-photon resonances and the profound hysteresis contrast with the case in Ref.~\cite{Camacho2021} where the high-density hysteresis branch respects the position of the multi-photon resonances. Therefore, one of the measurable consequences of the non-local interactions are shifts and broadenings of the lobular pattern.  In the regime where the non-local interactions are further suppressed with respect to the on-site interactions, for instance $V_{\text{od}}/U_X\approx 0.05$ one obtains a phase-diagram that closely follows the {\it bare and discrete} multi-photon resonances~\cite{Camacho2021}, as shown explictly in Fig.~\ref{FigA1} of Appendix.~\ref{appendixB}. Importantly, by means of the twist angle, one can therefore, enhance or suppress the effects of the non-local interactions on the optical response of the system~\cite{Julku2022}.

We also find that the non-local interactions lead to more metastable states, even within the space of population-balanced solutions. This multistability is somewhat reminiscent of the multistability seen in Fig.~\ref{Fig3} and Fig.~\ref{Fig5} and consists of steady state solutions with modified multi-photon resonance patterns. However we emphasise that it differs from the cases presented in Section.~\ref{hardcore} in that the populations remain balanced.
A set of four metastable solutions can be accessed through different hysteresis schemes which retain balanced populations, see Appendix~\ref{appendixB}.

For finite $U_X$ one can also obtain states with broken translational symmetry, such solutions require, however, smaller values of $\Omega/V_{nn}.$ In Fig.~\ref{Fig7} we show, for
$\Omega/V_{nn}=0.45$, solutions with unbalanced populations where red coloured exciton sites jump to a high density state with blue and green excitons smoothly increasing their density.  One recognizes the similarities between Fig.~\ref{Fig7} and the hard-core limit presented in Fig.~\ref{Fig3}(b) as both correspond to the same hysteresis protocol which give the same qualitative behaviour.
 \begin{figure}[ht!]
\centering
    \includegraphics[width=1\columnwidth]{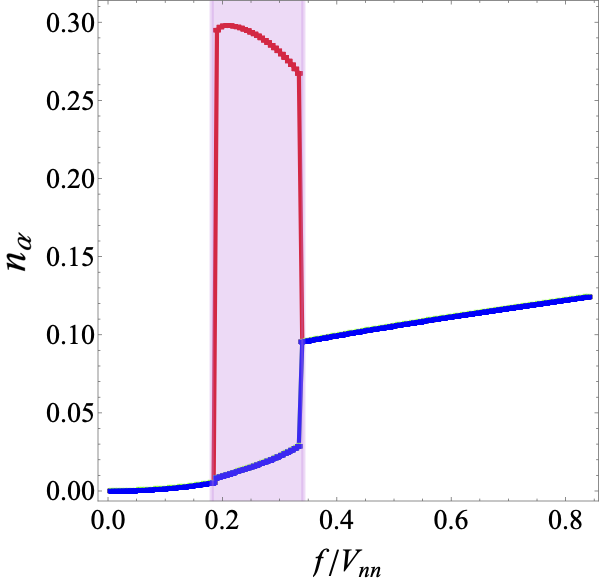}
    \caption{Exciton number per site for $\Omega/V_{nn}=0.45,$ and $V_{\text{od}}/U_X=0.2$, $\Omega/\gamma_c=\Omega/\gamma_x=1/3,$ $\Omega/\Delta\omega_c=-1/3$ and  $\Omega/\Delta\omega_X=2/3$. The pink area illustrates the regime with unbalanced populations.} 
    \label{Fig7}
\end{figure}
Experimentally, the branches in Fig.~\ref{Fig5} can be accessed through changing the direction of the drive $f,$ as commonly experimentally realized to detect bi-stabilities in conventional polaritons~\cite{Baas2004,Paraiso2010,Carusotto2013}. In general the states with complex hysteresis protocols are more challenging to access, as which of the states of differently broken translational symmetry will appear depends on how the translational symmetry breaking is seeded — by preparation or through underlying disorder.

\section{Experimental Perspectives and Conclusions}

\label{conclusions}
We have shown that non-local interactions $V_{nn}$ have significant qualitative effects on the non-linear optical response of exciton-polaritons in moiré materials. In addition to the unique moir\'e-induced non-linearities arising of the strong on-site interactions~\cite{Zhang2021,Camacho2021}, the non-local interactions reveal new features including steady states with broken translational symmetry, multi-stabilities, and deviations from the on-site multi-photon resonance conditions. To study these features, we  developed a self-consistent master equation based on a supercell containing three sites for excitons. These were treated independently at the mean-field level, allowing for the derivation of three coupled local master equations, which were solved self-consistently.

The predicted effects of the non-local interactions are readily measurable in experiment. They lead to a hysteretic dependence of the multi-photon resonance condition, wherein the form and position of the lobular pattern is determined by the direction of the drive (Fig.~\ref{Fig6}). Furthermore, a multi-valued hysteretic behaviour is found as a consequence of a spatial ordering of the excitons in the moir\'e sites.  The form of these steady states could be experimentally detected by spatially resolving the positions of the excitons~\cite{Lagoin2022}, but their existence is also apparent  in measurements of the  multi-stable hysteretic states of the cavity field~\cite{Baas2004,Paraiso2010} (see Fig.~\ref{Fig8}). 

Moir\'e systems are versatile platforms that allow for the control and manipulation of the excitonic properties over a wide range of parameters, for instance, by twisting the relative angle between the layers or by inserting an insulating layer between them to control the moir\'e superlattice properties and the features and degree of hybridisation of the excitons~\cite{Jiang2021}.  Our study encourages further studies to understand the behaviour of moir\'e excitons in different contexts. For instance, an intriguing avenue is to understand the interplay between multiple excitonic states, the strong exciton interactions, and the light-matter coupling. Another possibility is the study of the multi-photon lasing in the presence of non-local interactions~\cite{Camacho2021}. In addition, the role of free carriers has been demonstrated to be a powerful tool with which to control optical non-linearities in van der Waals heterostructures~\cite{Sidler2017,Julku2021,Tan2020,Emmanuele2020, Bastarrachea2021}. The formation of moir\'e trions has been recently experimentally reported~\cite{Liu2021,Shimazaki2020} and the nature of the optical signatures of trion-polaritons stands as an interesting open question~\cite{Camachosoon}.

Data supporting this publication are available in the Apollo repository~\cite{DataAccess}

\section{Acknowledgments}
We thank Atac Imamoglu for the careful reading of the manuscript and valuable comments. This work was partially supported by EPSRC Grant Nos. EP/P009565/1, EP/P034616/1 and by a Simons Investigator Award. ACG acknowledges grant No. IN108620 from DGAPA (UNAM). 

\appendix
\section{Dipolar interactions}
{We start discussing our approach for the non-local exciton-exciton interactions. Here, we assume that the non-local terms can be treated at the mean-field level, thus we have
\begin{gather}
\sum_{i\neq j} \frac{V_{ij}}{2}\hat x^\dagger_i\hat x^\dagger_j\hat x_j\hat x_i\rightarrow \sum_{i\neq j}
\frac{V_{ij}}{2}\langle \hat n_i\rangle\langle \hat n_j\rangle+
\sum_{i\neq j}V_{ij}\hat n_i \langle \hat n_j\rangle,
\label{mfdecouple}
\end{gather}
that is, we take $\langle (\hat n_i- \langle n_i\rangle)((\hat n_j- \langle n_j\rangle)\rangle\approx 0$. Only the last term in (\ref{mfdecouple}) will be relevant to the effective mean-field dynamics of the site $i$.

As illustrated in Fig.~\ref{FigAN} we introduce three exciton sites that colour the moir\'e lattice.
Thus, we define a larger supercell of three sites (labelled by $\alpha = R,G,B$) and denote the position of that  supercell by the index $I$ such that the site index $i \to (I,\alpha)$. In our mean-field ansatz the occupations of sites of the same colour $\alpha$ within all supercells are equivalent $\langle \hat{n}_{(I,\alpha)}\rangle = n_\alpha$.
\label{AppendixN}
 \begin{figure}[ht!]
\centering
    \includegraphics[width=1\columnwidth]{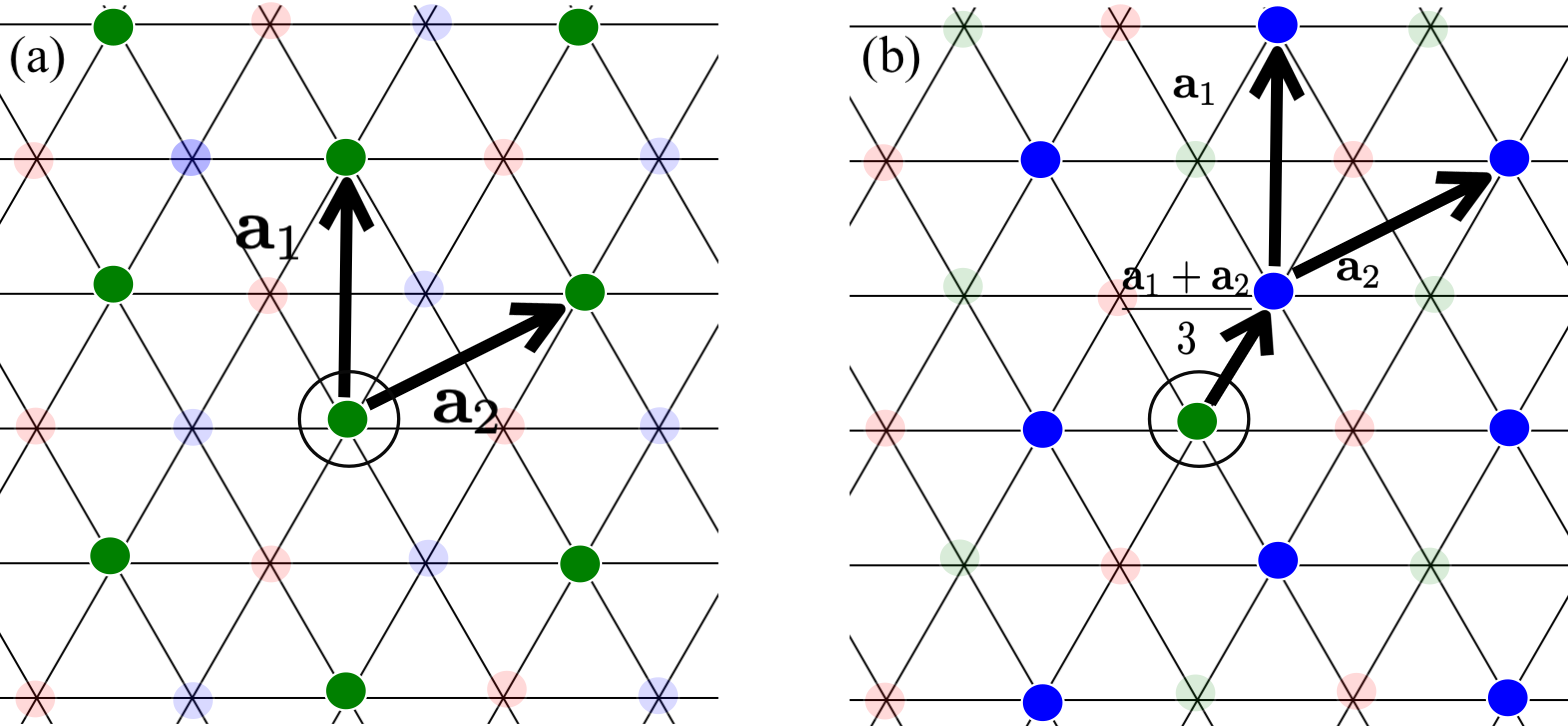}
    \caption{(a) Non-local interaction between one green site (circled) and  sites labelled with the same colour (b) Interaction between the green site (circled) and blue sites.} 
    \label{FigAN}
\end{figure}
Any given site  interacts with sites of the same and different colours. In Fig.~\ref{FigAN} we illustrate the non-local dipole-dipole interaction between a green site and (a) other green sites distanced by $\mathbf r^{gg}_{n,m}=|n\mathbf a_1+m\mathbf a_2|,$ from the original green site, (b) blue sites distanced by $\mathbf r^{gb}_{n,m}=|n\mathbf a_1+m\mathbf a_2+(\mathbf a_1+\mathbf a_2)/3|,$   where $n,m\in \mathbb{Z}.$ Here $\mathbf a_1=a_M(\sqrt{3},0)$ and $\mathbf a_2=a_M(3/2,\sqrt{3}/2) $ being $a_M$ the moir\'e lattice constant. [The red sites (not shown) are separated by $\mathbf r^{gr}_{n,m}=|n\mathbf a_1+m\mathbf a_2-(\mathbf a_1+\mathbf a_2)|/3$.

In detail, the interaction between a site $I$ with colour $\alpha$ and sites with the same colour is given by
\begin{gather}
\sum_{J\neq I} V_{(I,\alpha),(J,\alpha)}\hat n_{(I,\alpha)} \langle \hat n_{(J,\alpha)}\rangle = \sum_{J\neq I} V_{(I,\alpha),(J,\alpha)}\hat n_{(I,\alpha)}  n_{\alpha} \nonumber
\\ \nonumber
=n_\alpha\hat n_{I,\alpha}V_{nn}\sum'_{n,m}\frac{a_M^3}{|n\mathbf a_1+m\mathbf a_2|^{3}}\\
\approx n_\alpha\hat n_{(I,\alpha)}V_{nn}\times 2.12,
\end{gather}
where the sum in the second line is restricted to exclude $(n,m)=(0,0).$  Here, $V_{nn}=d^2/a_M^3.$ 

Similarly, the interaction between the site in supercell $I$ with colour $\alpha$ and all sites with a different colour $\beta\neq\alpha$ is 
\begin{gather}
\sum_{J}V_{(I,\alpha),(J,\beta)}\hat n_{(I,\alpha)} \langle \hat{n}_{(J,\beta)}\rangle 
= \sum_{J}V_{(I,\alpha),(J,\beta)}\hat n_{(I,\alpha)} n_\beta
\nonumber
\\
\nonumber = n_\beta\hat n_{(I,\alpha)}V_{nn}\sum_{n,m}\frac{a_M^3}{|n\mathbf a_1+m\mathbf a_2\pm\frac{\mathbf a_1+\mathbf a_2}{3}|^{3}} \\  
\approx n_\beta\hat n_{(I,\alpha)}V_{nn}\times 4.455,
\end{gather}
where the sign $\pm$ determines the colour of the sites. In Fig.~\ref{FigAN} the sign $\pm$ determines the coupling of the green sites to the blue ($+$) red ($-$) sites respectively,  both sums give  the same factor of $4.455;$ these sums are evaluated numerically and agree with the known results from Ref.~\cite{Lambin1993}.

Then, for the site $\alpha$ within any supercell we can define a local Hamiltonian that includes the non-local exciton-exciton interactions at the mean-field level,
\begin{gather}
  \left({V_{\rm d}}n_\alpha +{V_{\rm od}}\sum_{\beta\neq\alpha} n_{\beta}\right) \hat{n}_{(I,\alpha)}  
\end{gather}
with  $V_{\rm d} = 2.124\times V_{nn}$ and $V_{\rm od} = 4.455 \times V_{nn}$ as denoted in the main text. There, the number operator is written   $\hat{n}_{(I,\alpha)}= \hat{x}^\dag_\alpha\hat{x}_\alpha$, dropping the label $I$ of the supercell, since the mean-field self-consistency equation is the same for all supercells.

\section{Self-consistent scheme}
\label{appendixA}

We now provide the details of our self-consistent approach which is based on an iterative exact diagonalization of the three Linblad operators   $ \mathcal L_{\alpha}$ for $\alpha=\text{R,G}$ and $\text{B}$ that are coupled through the cavity field $\psi_\alpha$ and the long-range interaction term of the dipole-dipole interactions. The self-consistent approach consists of iteratively obtaining the exciton coherences $\langle \hat x_\alpha\rangle=x_\alpha$ and the populations $n_\alpha.$ 

The iterative scheme is obtained as follows and illustrated in Fig.~\ref{FigP} for a particular hysteresis branch.
\begin{enumerate}
    \item For a given $\Delta\omega_X$ we start from a large $f_0$ where the steady-state is single valued. Thus, we start from a random set of  parameters  $( x^0_\alpha(f_0),n^0_\alpha(f_0)).$ Here, the super-index denotes the step of the iteration which we now discuss. 
   \item We calculate  $\mathcal L_{\alpha}( x^0_\alpha(f_0),n^0_\alpha(f_0))$ and calculate the expectation values of the operators $\hat x_{\alpha}$ and $\hat n_{\alpha}$ which define the seed for the next iteration  $( x^1_\alpha(f_0),n^1_\alpha (f_0)),$ which interpolates $x^1_\alpha(f_0)=\langle \hat x_\alpha\rangle-\eta (\langle \hat x_\alpha\rangle-x^0_\alpha(f_0)),$ where $\eta$ is adjusted to speed numerical convergence.   Note that the sub-index has remained unchanged.
   \item We then iterate $\mathcal L_{\alpha}( x^i_\alpha(f_0),n^i_\alpha(f_0))$ using the parameters  $( x^{i-1}_\alpha(f_0),n^{i-1}_\alpha(f_0)).$ We iterate up to  $i=N_{max}=1800$ or when $\max(\text{error}_R,\text{error}_G,\text{error}_B)<10^{-8}$ where $\text{error}_\alpha=|n^{i-1}_\alpha-n^{i}_\alpha|$. The final state is denoted by  $( x^{N_{iter}}_\alpha(f_0),n^{N_{iter}}_\alpha(f_0)),$ where $N_{iter}$ is either $N_{max}$ or the number of iteration steps required to converge. Steps 2-3 are used for all of our numerics. 
   \item After convergence is achieved for $f_0,$ we then decrease $f_0$ by an amount of $\Delta f>0,$ we define $f_1=f_0-\Delta f.$  The initial seed  $( x^0_\alpha(f_1),n^0_\alpha(f_1))$ is no longer random and is taken as detailed below:

    \begin{itemize}
        \item Full hysteresis: 
        \begin{gather}  \nonumber
         x^0_\alpha(f_1)=x_\alpha^{N_{iter}}(f_0), \\ \nonumber
         n^0_\alpha(f_1)=n^{N_{iter}}_\alpha(f_0),
        \end{gather}
        for $\alpha\in\{\text{R,G,B}\}$ That is, all of the results obtained for $f_0$ are employed. This protocol is used for Fig.~\ref{Fig6}(c) and the high-density hysteresis branch in Fig.~\ref{Fig2}.
        \item Two excitons hysteresis: We retain only two solutions, that is, we take for instance
          \begin{gather}  \nonumber
         x^0_\alpha(f_1)=x_\alpha^{N_{iter}}(f_0), \\ \nonumber
         n^0_\alpha(f_1)=n_\alpha^{N_{iter}}(f_0), \\ \nonumber
         x^0_B(f_1)=0, \\ \nonumber
         n^0_B(f_1)=0,
        \end{gather}
        for $\alpha\in\{\text{R, G}\}$ .  We use this scheme in Fig.~\ref{Fig3}(a) and Fig.~\ref{Fig6}(c)
        \item Single exciton hysteresis. We retain only one solution, while the remaining needed parameters are set to zero. For instance, one of these branches corresponds to
        \begin{gather}
         x^0_R(f_1)=x_R^{N_{iter}}(f_0), \\ \nonumber
         n^0_R(f_1)=n_R^{N_{iter}}(f_0), \\ \nonumber
         x^0_G(f_1)=x^0_B(f_1)=0, \\ \nonumber
         n^0_G(f_1)=n^0_B(f_1)=0,
         \end{gather}
         we use this procedure of Fig.~\ref{Fig6}(b) and Fig.~\ref{Fig3}(b).  The procedure for this branch is illustrated in Fig.~\ref{FigP}.
        \item Lower branch: None of the solutions are kept, that is,
         $$( x^0_\alpha(f_1),n^0_\alpha(f_1))=(0,0),$$ for $\alpha\in \{\text{R,G,B}\}.$ This is the scheme followed for the lower branch in Fig.~\ref{Fig2} and Fig.~\ref{Fig6}(a).
    \end{itemize}
    \item We repeat step 2. Again, the arbitrary initial seed is only used to follow each hysteresis branch.
 
\end{enumerate}

\begin{figure}[ht!]
\centering
    \includegraphics[width=1\columnwidth]{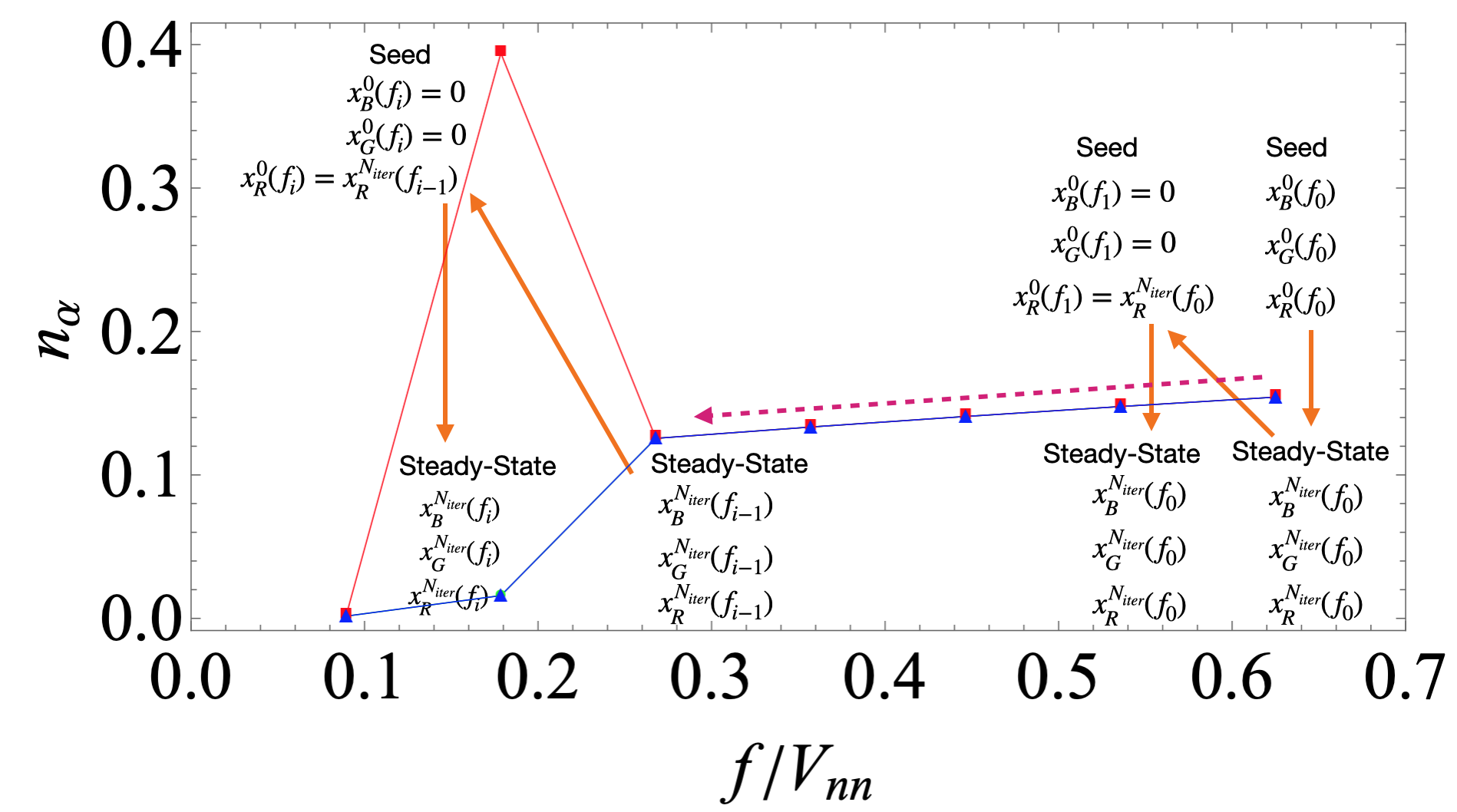}
    \caption{Cartoon of numerical protocol for the single exciton hysteresis. The hysteresis branch follows the pink dashed arrows, starting from a large value of $f_0$ where the steady-state is single valued. The seed for $f_1=f_0-\Delta f$ retains the steady-state outputs as illustrated in the figure and explained above. Thus, for $f_i$ with $i>0$ the seed breaks explicitly translational symmetry. (In this figure, for illustrative purposes,  we have greatly exaggerated the size $\Delta f$ which is kept much smaller in our analysis.)  } 
    \label{FigP}
\end{figure}

The convergence of our numerics is illustrated in Figs.~\ref{Fig2},~\ref{Fig3},~\ref{Fig8} and Fig.~\ref{Fig7} where error bars have been added and correspond to $|n^{N_{iter}}(f)-n^{N_{iter}-1}(f)|.$ The barely visible error bars confirm that our results are fully converged. For Sec.~\ref{hardcore}, the Hilbert space of the excitons is naturally restricted to having at most one exciton per-site. For finite on-site interactions we introduce a cut-off and restrict to ten excitons per site. The validity of this truncation depends on the strength of the drive and the exciton detuning that determine the exciton number. For the spanned parameters we restrict to exciton occupations much smaller than our cut-off and have indeed verified that our results do not change for larger sizes. Restricting to relatively small occupation number is also motivated by the experimental limitations to create arbitrary numbers of excitons per site through effects such as  saturation, population of higher bands, or even experimental damage of the samples due to the high intensity of the laser.

\section{Hysteresis schemes}
\label{appendixB}
The different hysteresis schemes permit multi-stabilities. As mentioned in the main text, the low- and high-density hysteresis branches in Fig.\ref{Fig5} are accompanied by two additional branches that can be accessed via the mechanisms discussed in the Appendix~\ref{appendixA}. These solutions are illustrated in Fig.\ref{Fig6}

\begin{figure*}[ht!]
\centering
    \includegraphics[width=1.92\columnwidth]{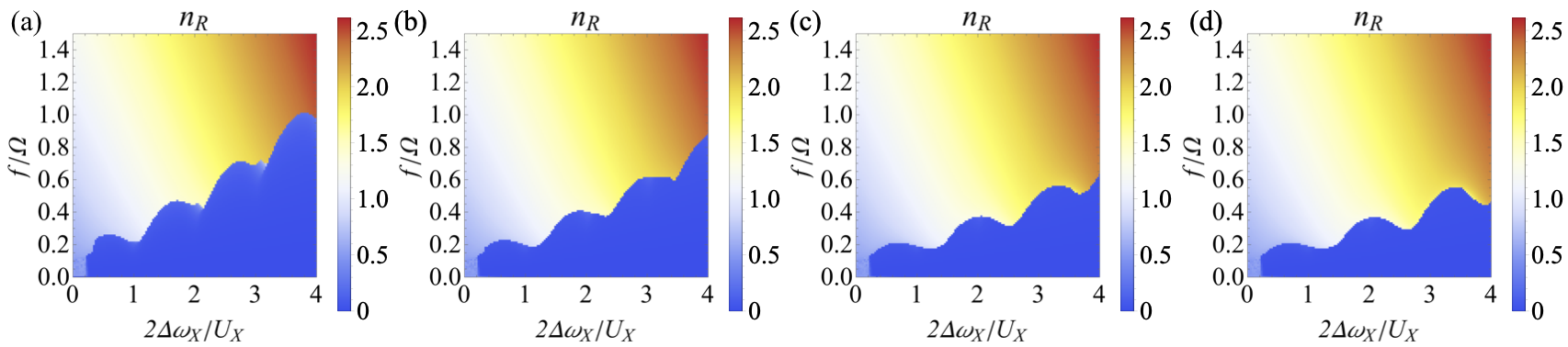}
    \caption{Exciton number per site for steady states with balanced colour population. (a) Low-density branch. (b) One exciton hysteresis protocol, (c) Two exciton hysteresis protocol and (d) full hysteresis scheme as explained above. } 
    \label{Fig6}
\end{figure*}
The steady states in Figs.~\ref{Fig6} do not break translational symmetry, however, the stark difference between the phase-diagrams strongly depend on the initial seed of our numerics. Figure.~\ref{Fig6}(b) corresponds to the high-density hysteresis branch for a single coloured exciton. 

Figure.~\ref{Fig6}(c) corresponds to the hysteresis where two coloured excitons are recursively iterated.  In addition, the deformation of the multi-photon resonances becomes more visible.
 
Finally, in Fig.~\ref{FigA1} we set a much smaller value of the non-local interactions $V_{\text{od}}/U_X$ to demonstrate that the {\it local}  multi-photon resonances of Eq.~\ref{Nres-as-ratio} are recovered in this limit. 
 
 \begin{figure}[ht!]
\centering
    \includegraphics[width=.92\columnwidth]{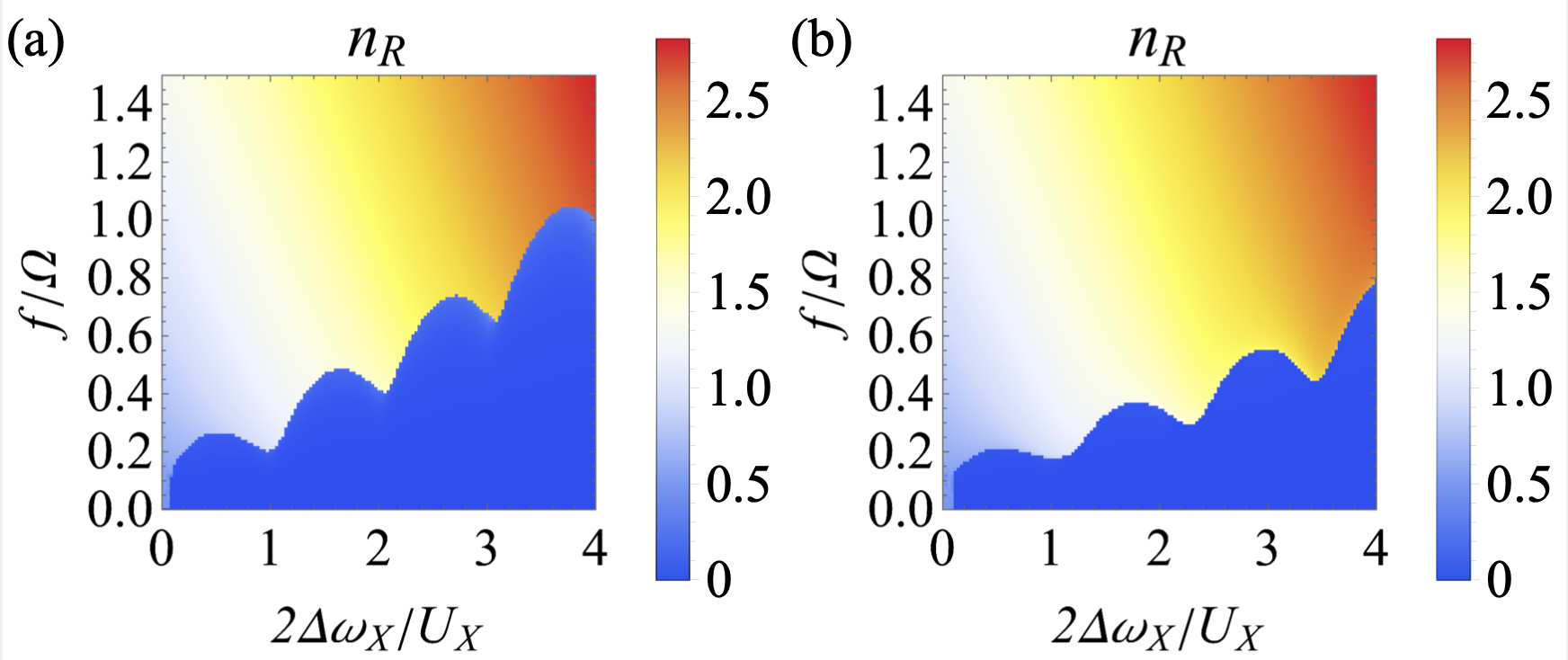}
    \caption{Exciton number per site for steady states with balanced colour population. Here we illustrate the (a) low-density and (b) full hysteresis for $V_{\text{od}}/U_X=0.05$ and remaining same parameters as in Fig.~\ref{Fig6} } 
    \label{FigA1}
\end{figure}
\bibliography{moire}
\end{document}